# Capturing 3D atomic defects and phonon localization at the 2D heterostructure interface

Xuezeng Tian[1,2†], Xingxu Yan[3,4†], Georgios Varnavides[5,6,7†], Yakun Yuan[1], Dennis S. Kim[1], Christopher J. Ciccarino[5,8], Polina Anikeeva[6,7], Ming-Yang Li[9], Lain-Jong Li[10], Prineha Narang[5*], Xiaoqing Pan[3,4,11*], Jianwei Miao[1*]

[1]*Department of Physics & Astronomy and California NanoSystems Institute, University of California, Los Angeles, CA 90095, USA.* [2]*Beijing National Laboratory for Condensed Matter Physics, Institute of Physics, Chinese Academy of Sciences, Beijing, 100190, China.* [3]*Department of Materials Science and Engineering, University of California - Irvine, Irvine, California 92697, USA.* [4]*Irvine Materials Research Institute, University of California - Irvine, California 92697, USA.* [5]*Harvard John A. Paulson School of Engineering and Applied Sciences, Harvard University, Cambridge, MA 02138, USA.* [6]*Department of Materials Science and Engineering, Massachusetts Institute of Technology, Cambridge, MA 02139, USA.* [7]*Research Laboratory of Electronics, Massachusetts Institute of Technology, Cambridge, MA 02139, USA.* [8]*Department of Chemistry and Chemical Biology, Harvard University, Cambridge, Massachusetts 02138, USA.* [9]*Physical Sciences and Engineering Division, King Abdullah, University of Science and Technology, Thuwal, 23955-6900, Kingdom of Saudi Arabia.* [10]*Department of Mechanical Engineering, The University of Hong Kong, Pokfulam Road, Hong Kong .* [11]*Department of Physics and Astronomy, University of California - Irvine, Irvine, California 92697, USA.*

[†]These authors contributed equally to this work. [*]Correspondence author. Email: miao@physics.ucla.edu (J.M.); xiaoqinp@uci.edu (X.P.); prineha@seas.harvard.edu (P.N.)

**Abstract:** The 3D local atomic structures and crystal defects at the interfaces of heterostructures control their electronic, magnetic, optical, catalytic and topological quantum properties, but have thus far eluded any direct experimental determination. Here we determine the 3D local atomic positions at the interface of a $MoS_2$-$WSe_2$ heterojunction with picometer precision and correlate 3D atomic defects with localized vibrational properties at the epitaxial interface. We observe point defects, bond distortion, atomic-scale ripples and measure the full 3D strain tensor at the heterointerface. By using the experimental 3D atomic coordinates as direct input to first principles calculations, we reveal new phonon modes localized at the interface, which are corroborated by spatially resolved electron energy-loss spectroscopy. We expect that this work will open the door to correlate structure-property relationships of a wide range of heterostructure interfaces at the single-atom level.

**Main Text**: A major challenge in materials design and engineering is to tailor the 3D atomic structures at the interface to achieve the desired properties. Although 2D lateral and vertical heterostructures have been actively studied for fundamental interest and practical applications (*1-9*), our current understanding of the atomic structure at the heterointerface has been primarily relied on aberration-corrected electron microscopy and scanning probe microscopy (*6-11*), which provide either 2D projection images or surface structure. On the computational side, density functional theory (DFT) can be used to predict the properties of heterostructures (*12-15*), but requires knowledge of the 3D local atomic coordinates. Due to the difficulty in directly measuring these 3D coordinates, such studies often use perfect crystal lattices (*12-14*), statistically incorporate crystal defects into the interface (*15*), and relax the atomic configurations to the minimum energy states. However, real heterointerfaces have neither perfect crystal lattices, nor are in the minimum energy states. Here, using a $MoS_2$-$WSe_2$ lateral heterojunction as a model, we applied atomic electron tomography (AET) (*16-18*) to determine, for the first time, the 3D atomic coordinates and crystal defects at the heterointerface with picometer precision. We observed various crystal defects, including vacancies, substitutional defects, bond distortion and atomic-scale ripples, and quantitatively characterized the 3D atomic displacements and full strain tensor across the heterointerface. The experimentally measured 3D atomic coordinates, representing a metastable state of the heterojunction, were used as direct input to first principles calculations to reveal new phonon modes localized at the heterointerface, which were corroborated by the measurements of spatially resolved electron energy-loss spectroscopy (EELS). In contrast, the phonon dispersion derived from the minimum energy state of the heterojunction is absent of the local interface phonon modes, indicating the importance of using experimental 3D atomic coordinates as direct input to better predict the properties of heterointerfaces.

The experiment was conducted with an aberration-corrected scanning transmission electron microscope (STEM), operated at 60 kV in annular dark-field (ADF) mode. A tilt series of 12 images was acquired from an interface region of an epitaxial $MoS_2$-$WSe_2$ lateral heterojunction (see supplementary materials, Fig. 1A, fig. S1, tables S1 and S2). Using scanning AET (*19*), we determined the 3D atomic coordinates and atomic species of an interface region, containing 488 Mo, 991 S, 150 W, 257 Se atoms and 16 S/Se vacancies (Fig. 1B, C, fig. S2 and table S1). From the experimental 3D atomic model, we combined multislice simulations with scanning AET to estimate the 3D precision of the method to be 4 pm, 15 pm, 6 pm and 15 pm for the Mo, S, W and Se atoms, respectively (see supplementary materials). We found that there are 38 S atoms and 8 Se atoms at the interface, indicating that during the two-step epitaxial growth, the first grown $WSe_2$ layer had W-terminals and favored W-S bonds (see supplementary materials). A majority of the S and Se vacancies (9 out of 16) are located along the interface, suggesting the 2D interface lowers the vacancy formation energy. We also observed S/Se substitutional defects close the interface (Fig. 1C).

Since the lattice constant of WSe$_2$ is 5.53% larger than that of MoS$_2$ (6), the lattice mismatch must be accommodated by significant crystal defects at the interface. Figure 1D shows atomic-scale ripples at the heterostructure interface. A side view perpendicular to the interface shows a sinusoidal oscillation of the ripples with an amplitude of ~1.5 Å in the z-axis and a wavelength of ~5 nm along the interface (Fig. 1E). We divided the reconstructed region into nine strips along the interface and calculated the standard deviation (roughness) of the z-coordinates within each strip. The roughness is largest on the WSe$_2$ side of the interface, propagating into the MoS$_2$ side, and decaying away from the interface (Fig. 1F). The atomic-scale ripples induced bond distortion at the interface, which we quantified by measuring the bond angles between the S/Se and Mo/W atoms. As the interface in the reconstructed region mainly falls in the valley of the ripple (Figs. 1D and 2A), the bottom layer has larger bond angles than the top layer and the average difference is 7.0° (Fig. 2B).

The lattice mismatch also creates atomic displacements away from the perfect crystal lattice (Fig. 2C). From the displacements of all atoms, we determined the full 3D strain tensor at the heterointerface. Fig. 2D and fig. S3 show the six components of the strain tensor in the three atomic layers. The in-plane components $\varepsilon_{xx}$ and $\varepsilon_{yy}$ are positive on the MoS$_2$ side and negative on the WSe$_2$ side, while the shear component $\varepsilon_{xy}$ is negative along the interface. This is consistent with the fact that the smaller lattice constant of MoS$_2$ than of WSe$_2$ induces an expansion in MoS$_2$ and a compression in WSe$_2$ to achieve an epitaxial interface. We also observed that $\varepsilon_{zz}$ is the largest among all six components, which is due to the lack of constraint along the z-axis, and consistent with observations of the out-of-plane ripple. Although strain engineering in 2D heterointerfaces has been previously reported (*9, 20*), our results represent the first experimental measurements of the 3D strain tensor in a 2D heterostructure with near atomic resolution, showing the out-of-plane component $\varepsilon_{zz}$ is at least as important as the other ones.

Next, we investigated the vibrational properties at the heterointerface by utilizing the experimental 3D atomic coordinates as direct input to first principles calculations without relaxation. Two atomic strips perpendicular to interface were chosen and stitched together to create an interface supercell, consisting of 144 atoms with periodic boundary conditions (see supplementary materials, Fig. 3A and C). The phonon dispersion was calculated for the interface supercell, which was unfolded onto an effective band structure with the symmetry of the primitive hexagonal cell (Fig. 3B and fig. S4) (*21*). The colorscale represents the supercell spectral function with higher intensity arising from bulk $MoS_2$ and $WSe_2$ (*22*). The presence of lower-intensity, dispersion-less modes such as those labelled by the polygon markers, suggests the emergence of localized, interface modes. There are also shadow bands in the spectral function, introduced primarily by the 3D atomic defects at the heterointerface (*19*). Investigation of the phonon eigen-displacements at the Γ, M, and K points indeed reveals localized interface modes (Fig. 3C), which are not present in the first principles calculations of the minimum energy state of the heterojunction (fig. S4D). To quantitatively characterize the phonon dispersion, we used the least-squares fit of a linear combination of the bulk $MoS_2$ and $WSe_2$ phonon density of states (PDOS) (Fig. 3D bottom). Three distinctive peaks were observed in the PDOS of the interface dispersion, namely 17.5, 28.5, and 41.5 meV, but are absent in the least-squares fit PDOS (Fig. 3D top). To further analyze the phonon localization, we computed the participation ratios (PR), a measure of the deviation away from plane-wave like phonon vibrations expected in perfect crystal lattices (*23*). Figure 3E shows the PR as a function of phonon frequency. The phonon modes with PR smaller than unity illustrate the disordered nature introduced by the 3D atomic defects at the interface (top). We correlated this localization with the interface by computing an interface-weighted PR (see supplementary materials). Increased density of phonon modes with interface-weighted PR above average (dashed line), labeled with a triangle, square and pentagon in Fig. 3E (bottom), further confirm the existence of interface-localized phonon modes (Figure 3B, 3C).

To experimentally corroborate the interface phonon modes, we measured the vibrational spectra of the $MoS_2$–$WSe_2$ heterostructure using spatially resolved EELS with an energy resolution of 5.7 meV at 60 keV (see supplementary materials) (*24, 25*). By scanning continuously from a $MoS_2$ to a $WSe_2$ region (Fig. 4A and B), we collected a series of vibrational spectra across the heterointerface (Fig. 4C), which reflect the PDOS throughout the entire Brillouin zone (*26, 27*). Based on different features at the $MoS_2$, interface, and $WSe_2$ regions, the vibrational spectra were divided into three groups via k-means clustering (fig. S6). To enhance the signal-to-noise ratio, the spectra in each group were summed up to produce an average spectrum, of which the peak positions were fit with multiple Gaussians (Fig. 4D, fig. S7 and table S3). We found that the peak positions of the average spectra in the $MoS_2$ and $WSe_2$ regions agree with those of previously reported PDOS curves for bulk $MoS_2$ and $WSe_2$ (table S3). However, the average interface spectrum shows two peaks at 27.9 meV and 41.1 meV, which are not present in the PDOS curves of bulk $MoS_2$ and $WSe_2$. Both peaks are consistent with the second and third interface modes of 28.5 meV and 41.5 meV, obtained by DFT calculations with the experimental 3D atomic coordinates (Fig. 3). The first interface mode of 17.5 meV in the DFT calculations could not be detected in the EELS spectrum due to the influence of the zero-loss peak tail.

The observation of localized interface phonon modes has implications on the thermal transport mechanism across the interface. When phonons propagate across an interface of two materials with different vibrational properties, phonons undergo additional scattering which impedes heat transport across the interface (*28-30*). In the case of the MoS$_2$-WSe$_2$ heterojunction, the difference in PDOS, suggests several optical phonon modes in WSe$_2$ cannot scatter into MoS$_2$ while simultaneously conserving their energy and momentum. What is more, heat is predominantly carried through large group-velocity plane-wave like lattice vibrations. As our first principles calculations suggest, the emergence of localized modes at the interface will lead to increased scattering and thus an increased thermal interface resistance across the heterostructure. Furthermore, our results show that the interface phonon modes cannot be obtained by a linear combination of the bulk phonon modes of two different materials forming the interface, but rather depend on the 3D local atomic structure and crystal defects at the interface, including point defects, bond distortions, atomic-scale ripples and strain. Thus, atomic-scale design and engineering of interface phonon modes could potentially improve the thermal transport in 2D and 3D heterostructures, mitigate the heat dissipation, and increase the lifetime of modern microelectronic devices.

Using a monolayer MoS$_2$-WSe$_2$ heterojunction as a model, we demonstrated a correlative experimental and first principles method to determine the 3D atomic coordinates and crystal defects and capture the localized vibrational properties at the epitaxial interface. We observed the local atomic structure reconstruction of the interface, by means of out-of-plane ripples and bond distortion to accommodate the lattice mismatch. We measured the full 3D strain tensor at the interface, showing $\varepsilon_{zz}$ is the largest among the six components. The experimentally measured 3D atomic coordinates, representing a metastable state of the heterojunction, were directly used in first principles calculations of vibrational properties. The unfolded phonon dispersion relation and the PDOS show distinctive phonon modes emerging at the interface, which were verified by spatially resolved EELS. Although in this study we revealed 3D atomic defects and local interface phonon modes in a $MoS_2$-$WSe_2$ heterojunction, the correlative method could be used to probe the physical, material, chemical and electronic properties of various heterointerfaces at the single-atom level. Looking forward, we expect that the ability to couple the 3D atomic structures and crystal defects with the properties of heterostructure interfaces will transform materials design and engineering across different disciplines.

15. S. Thomas, M.A. Zaeem, *Phys. Chem. Chem. Phys*. **22**, 22066-22077 (2020).

16. M. C. Scott *et al*., *Nature* **483**, 444-447 (2012).

17. J. Miao, P. Ercius, S. J. L. Billinge, *Science* **353,** aaf2157 (2016).

18. Y. Yang *et al., Nature* **542,** 75–79 (2017).

19. X. Tian *et al., Nat. Mater.* **19**, 867-873 (2020).

20. C. Zhang *et al*., *Nat. Nanotechnol.* **13**, 152–158 (2018).

21. Y. Ikeda *et al.*, *Phys. Rev. B* **95**, 024305 (2017).

22. C. Ciccarino *et al, Nano Lett.* **18**, 5709–5715 (2018).

23. T. Tadano, Y. Gohda, S. Tsuneyuki, *Phys. Rev. Lett*. **114**, 095501 (2015).

24. X. Yan *et al*., *Nano Lett.* **19**, 7494–7502 (2019).

25. X. Yan *et al*., *Nature* **589**, 65–69 (2021).

26. O. L. Krivanek *et al*., *Nature*. **514**, 209–212 (2014).

27. K. Venkatraman *et al*., *Nat. Phys.* **15**, 1237–1241 (2019).

28. Y. Ni *et al*., *Int. J. Heat Mass Transf.* **144**, 118608 (2019).

29. R. M. Costescu, M. A. Wall, D. G. Cahill, *Phys. Rev. B* **67**, 054302 (2003).

30. E. T. Swartz, R. O. Pohl, *Rev. Mod. Phys*. **61**, 605 (1989).
**ACKNOWLEDGMENTS**: **Funding**: This work was primarily supported by the US Department of Energy (DOE), Office of Science, Basic Energy Sciences, Division of Materials Sciences and Engineering under award DE-SC0010378 and DESC0014430. We also acknowledge support by the Army Research Office MURI program under grant no. W911NF-18-1-0431, STROBE: a National Science Foundation (NSF) Science and Technology Center under award DMR1548924, and the NSF DMREF program under award DMR-1437263. The work at UC Irvine was partially supported by the NSF through the University of California-Irvine Materials Research Science and9

Engineering Center under award DMR-2011967. TEM experiments were conducted using the facilities in the Irvine Materials Research Institute (IMRI) at the University of California, Irvine.

**Author contributions**: J.M. directed the project; X.Y., X.T., X.P. and J.M. initially designed and/or conducted the experiments; M.Y. L. and L.-J. L. synthesized the sample; X.T., Y.Y., D.K. and J.M. performed 3D reconstruction, atom tracing and classification, and analyzed the data; G.V., C.J.C., P.A. and P.N. carried out first principles calculations; J.M., X.T., X.Y., and G.V. wrote the manuscript. All authors commented on the manuscript. **Competing interests**: The authors declare no competing financial interests. **Data and materials availability**: All data are available in the main text or the supplementary materials.

**SUPPLEMENTARY MATERIALS**

Materials and Methods

Figures S1-S7

Tables S1 and S3

References (*31-46*)

**FIGURES**

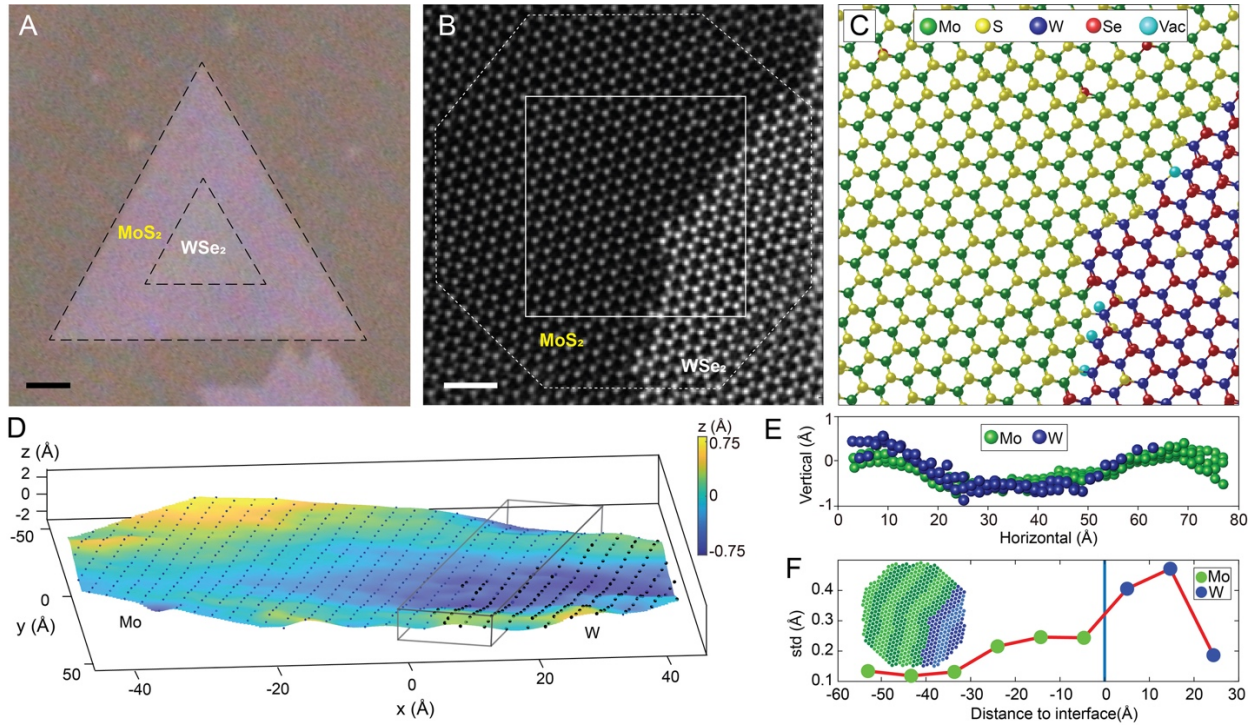

**Fig. 1. 3D atomic coordinates of a monolayer MoS$_2$-WSe$_2$ heterostructure.** (**A**) An optical image of the as grown sample on a sapphire substrate. Scalebar, 2 um. (**B**) Top view of the 3D reconstruction of an interface region in the heterostructure. The 3D atomic coordinates and species in the octagon were determined with picometer precision. Scalebar, 1 nm. (**C**) Experimental 3D atomic model of the square region in (B). (**D**) Atomic-scale ripples in the interface region, where the dots represent the Mo/W atoms. The gray boxes indicate the interface region with large ripples. (**E**) Side view of the gray box region in (D), showing a sinusoidal oscillation at the interface. (**F**) Distribution of the roughness, i.e. the standard deviation of the z-coordinates, within each of the nine strips (inset) as a function of the distance to the interface.

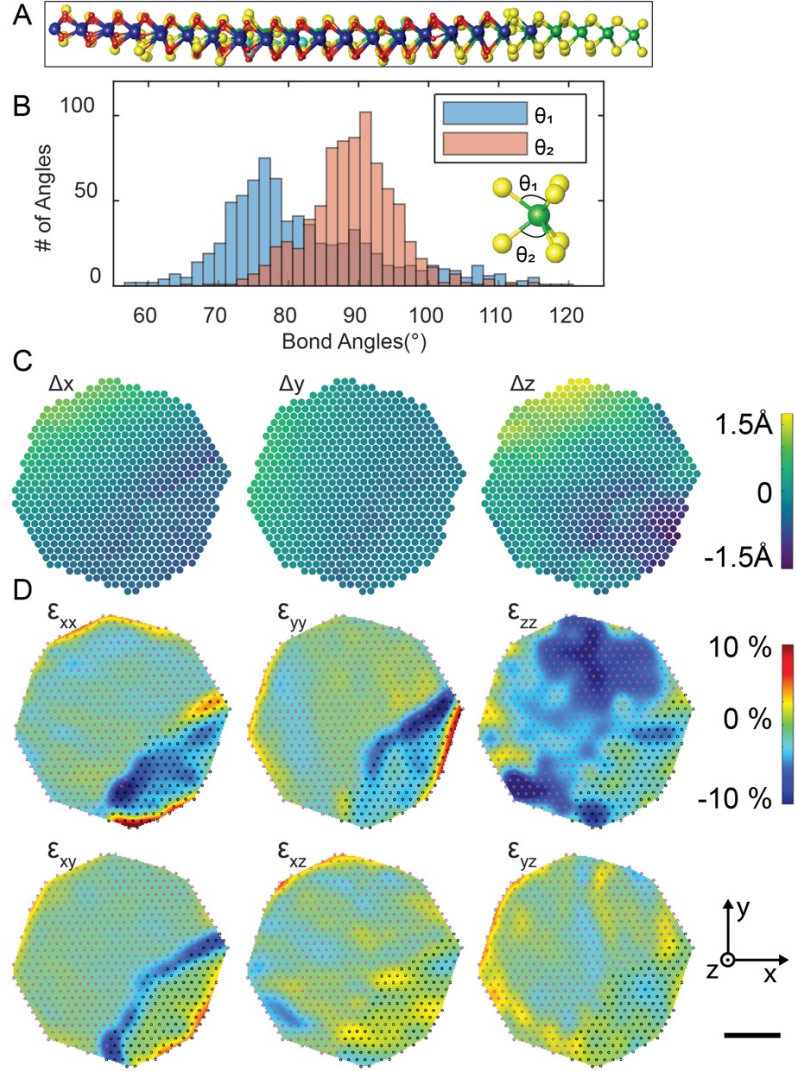

**Fig. 2. 3D atomic displacements and strain tensor at the 2D heterostructure interface.** (**A**) Side view of the bond distortion along the $MoS_2$-$WSe_2$ interface. (**B**) Distribution of the bond angles between the S/Se and Mo/W atoms. The average bond angle in the bottom layer is 7.0° larger than that on the top layer. **C**. 3D atomic displacements of Mo/W atoms along the x-, y- and z-axis. **D**, Six components of the strain tensor in the Mo (weak dots)/W(dark dots) layer. The strain tensor of the two S/Se layers is shown in fig. S3.

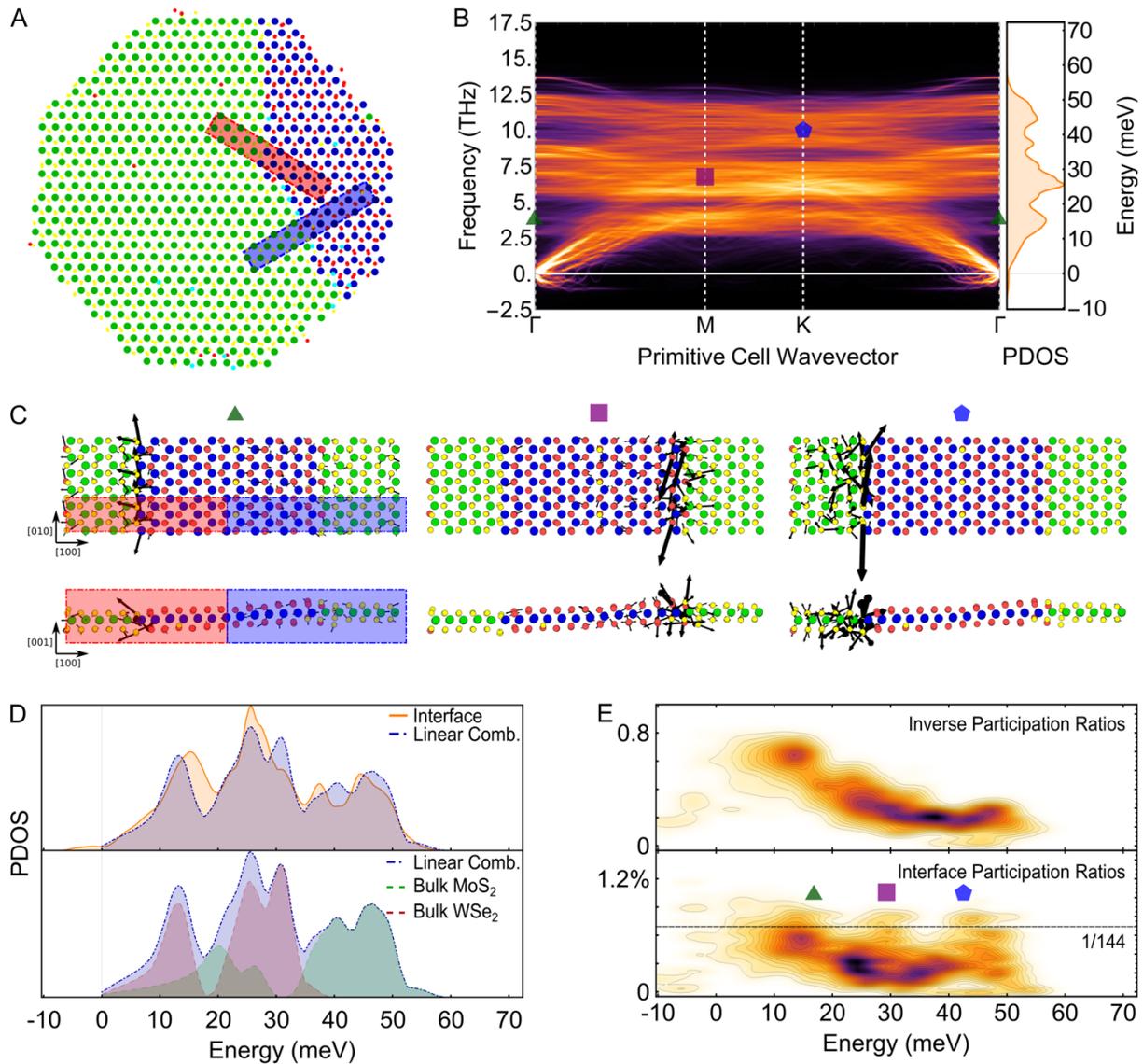

**Fig. 3. Phonon dispersion and localized interface modes.** (**A**) Top view of the experimental 3D atomic model, where two atomic strips (red and blue) perpendicular to interface were stitched together to create an interface supercell for DFT calculations. (**B**) Phonon dispersion calculated from the interface supercell and unfolded onto a hexagonal primitive cell. (**C**) Eigen-displacements of three interface modes at high symmetry points with increasing frequencies, labeled with a triangle, square and pentagon in (B). Red and blue strips represent the stitched interface supercell and the displacement arrows are magnified by 10x for clarity. (**D**) Orange and blue curves represent

the PDOS obtained from (B) (top) and by the least-squares fit of a linear combination of the bulk MoS$_2$ and WSe$_2$ dispersion (bottom), respectively. (**E**) Inverse participation ratios for the interface supercell (top) and weighted by the distance to the interface (bottom). In the latter, the emergence of high interface participation contributions at the three energy windows (labeled with a triangle, square and pentagon) becomes apparent with intensities larger than the average (1/144, where 144 is the number of atoms in the stitched supercell.)

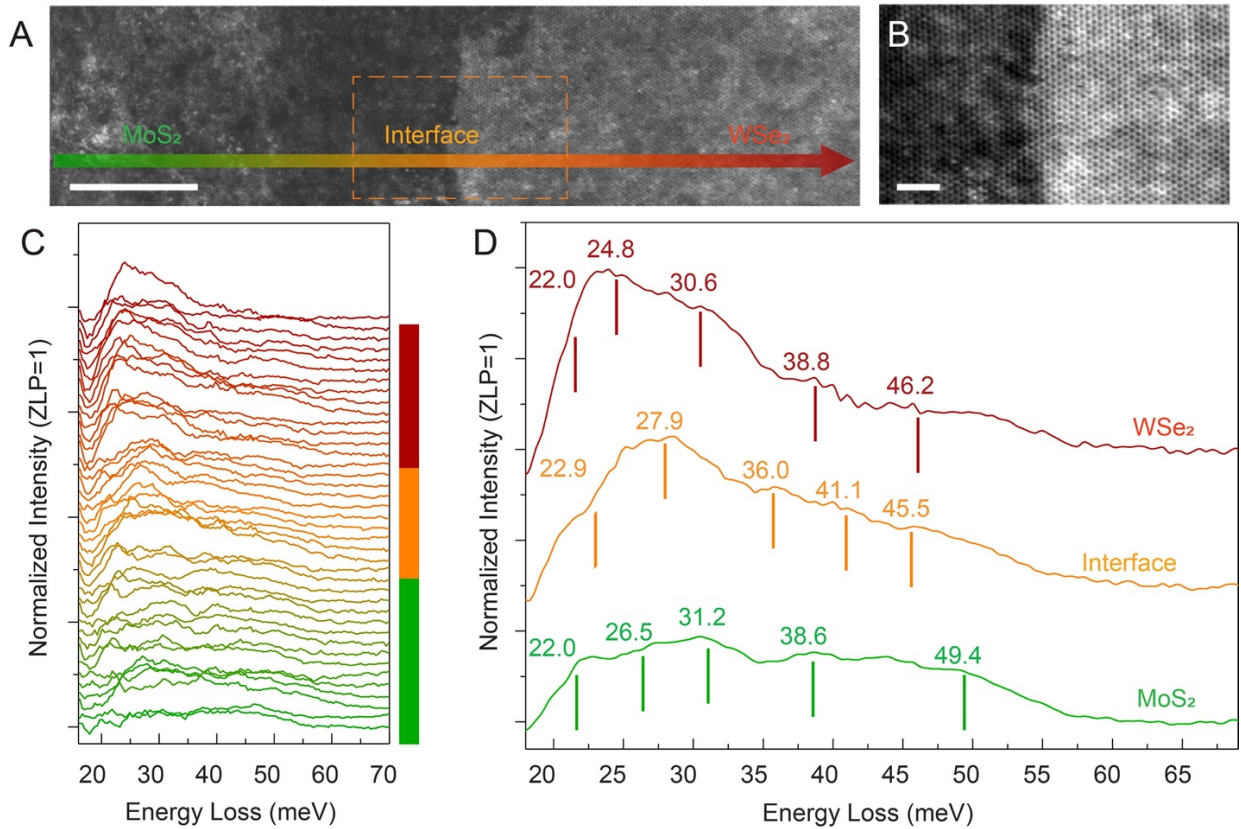

**Fig. 4. Vibrational spectra of the MoS$_2$–WSe$_2$ heterointerface.** (**A**) Low magnification ADF-STEM image of the heterointerface. Scale bar, 10 nm. (**B**) Magnified ADF-STEM image of the orange box in (A), showing the local atomic structure at the interface. Scale bar, 2 nm. (**C**) Vibrational spectra measured from the MoS$_2$ region (green) through the interface (orange) to the WSe$_2$ region (brown) with a step size of 1.6 nm. The background was subtracted by fitting a power function to each spectrum (fig. S5). K-means clustering was used to separate the spectra in the three regions (see supplementary materials). **(D)** Average vibrational spectra of the $WSe_2$, interface, and $MoS_2$ region, where the peak positions were identified by using multi-Gaussian fit (fig. S7).

Supplementary Materials for

**Capturing 3D atomic defects and phonon localization at the 2D heterostructure interface**

Xuezeng Tian[1,2†], Xingxu Yan[3,4†], Georgios Varnavides[5,6,7†], Yakun Yuan[1], Dennis S. Kim[1], Christopher J. Ciccarino[5,8], Polina Anikeeva[6,7], Ming-Yang Li[9], Lain-Jong Li[10], Prineha Narang[5*], Xiaoqing Pan[3,4,11*], Jianwei Miao[1*]

[†]These authors contributed equally to this work. [*]Correspondence author.

**This PDF file includes:**

Materials and Methods
Figs. S1 to S7
Tables S1 to S3
References

**Materials and Methods**

**Sample Preparation**

Single crystal WSe$_2$ monolayer was first grown by the chemical vapor deposition method. WO$_3$ power of 0.6 g was placed in a quartz boat located in the heating zone center of the furnace. A sapphire substrate was placed at the downstream side, just next to quartz boat. The Se powders was placed in a separate quartz boat at the upper stream side of the furnace and the local temperature was maintained at 260 °C during the reaction. The gas flow was brought by a mixture of Ar (90 sccm) and H$_2$ (6 sccm) with a chamber pressure of 20 Torr. After reaching the required growth temperature 925 °C, the heating was kept for 15min and the furnace was then naturally cooled down to room temperature. After optical characterizations for the as-grown WSe2, the sample was then put into a separate furnace for the second step of MoS$_2$ growth. The setup for MoS$_2$ growth is similar to that of WSe$_2$, by switching the source to MoO$_3$ power (0.6 g) and S powers. The Ar gas flow was set at 70 sccm and the pressure was controlled at 40 Torr. The sapphire substrate with WSe$_2$ sample was placed at the downstream side of MoO$_3$ boat and the distance between sample and quartz boat was 9 cm to achieve the best Mo and S sources ratio to form the WSe$_2$/MoS$_2$ heterojunction. The center zone and S source region were heated to 755 °C and 190 °C, respectively, and hold 15 min for synthesis, and then naturally cooled down to room temperature. The as-grown WSe$_2$-MoS$_2$ heterojunction was then transferred onto a QUANTIFOIL® holey carbon film TEM grid by a poly (methyl methacrylate) (PMMA) (950 PMMA A4, Micro Chem) assisted transfer method. PMMA thin film was spin-coated on top of sample, and then the PMMA/sample/sapphire was dipped in a 6M HF solution to etch the sapphire substrate. PMMA/sample was lifted from the etching solution and diluted and cleaned in DI water, and then transferred onto TEM grid. The PMMA layer was rinsed out by acetone and isopropanol. The as-prepared TEM grid was vacuum annealed at 160 °C for over 8 hours to remove most of the residual polymer before loading into the electron microscope.

**Tomographic data acquisition of a MoS$_2$–WSe$_2$ heterojunction**

The AET experiment was performed on a MoS$_2$–WSe$_2$ heterojunction using a Nion UltraSTEM 200 equipped with C3/C5 corrector and high-energy resolution monochromated EELS system. ADF-STEM images were acquired at 60 kV with an averaged beam current of 17 pA and a probe size of 1.3 Å. The convergence semi-angle of electron probe was 38 mrad and the inner and outer collection semi-angles of ADF detector were 50 and 210 mrad, respectively, where a small detector inner angle was chosen to reduce the electron damage to the sample. Table S1 summarizes the experimental parameters. A tilt series of 12 angle were acquired from the heterojunction (table S2). To reduce the total electron dose and eliminate the sample drift issue, 10 images per tilt angle were collected with a dwell time of 4 μs per pixel and 512×512 pixels for each image. Due to the use of a low electron energy beam and a low-dose acquisition scheme, there was minimal structural change of the heterojunction during the data acquisition (fig. S1). The total electron dose for the whole tilt series was estimated to be $3.8×10^5$ e/Å$^2$.

**Image processing**

Drift correction, denoising and background subtraction. Drift correction was applied to each image as described elsewhere (*18, 19, 31*). At each tilt angle, 10 images were registered and the remaining 9 images were aligned to the first image with cross correlation to sub-pixel precision. The drift between two neighbouring images was typically less than one pixel and the drift vector followed a linear trajectory. As the drift vector was uniformly distributed along the slow scan direction during data acquisition, the scan distortion was implemented by interpolating the non-corrected image onto the drift corrected pixel positions. After the drift correction, the 10 images were summed to obtain an average image for the tilt angle. As each image contains mixed Poisson and Gaussian noise, a block-matching and 3D filtering algorithm was applied to denoise the average image of each tilt angle (*32*). From the denoised image, the background was estimated based on the fact that the center of the hexagonal atomic lattice should be empty. The local minimum of each hexagonal lattice was chosen to form a sparse map, and the background was obtained by interpolating the local minimum map with a radial basis function (*33*). The estimated background was subtracted from the denoised image.

Angle calibration. The nominal tilt angles were recorded during data acquisition (table S2) but must be calibrated due to the hysteresis of the stage. For each image, the x and y coordinates of the W and Mo atoms were determined by fitting a 2D Gaussian. To calibrate the tilt angles, a least square method was used to minimize $E$ (*19*), defined as

$$E = \sum_i \sum_j \left\{ \left[ P_x(\mathbf{r}_i, \alpha_j, \beta_j) - x_i^j \right]^2 + \left[ P_y(\mathbf{r}_i, \alpha_j, \beta_j) - y_i^j \right]^2 \right\}, \qquad (1)$$

where $P_x$ and $P_y$ are the functions of projecting the 3D coordinates of the W and Mo atoms to the x and y coordinates in the image, respectively, $\mathbf{r}_i$ is the 3D coordinates of the $i^{th}$ W or Mo atom, $\alpha_j$ and $\beta_j$ are the tilt angles of the $j^{th}$ projection, $x_i^j$ and $y_i^j$ are the measured x and y coordinates in the $i^{th}$ atom in the $j^{th}$ projection, respectively. By minimizing $E$, we calibrated the tilt angles of the 12 images (table S2).

Vibration correction. Due to the free-standing nature of the monolayer 2D heterojunction during data acquisition, we observed that the 2D heterojunction vibrated in the vertical direction. As a result, the atoms in the ADF-STEM images are elongated and blurred at relatively high tilt angle. The vibration blurring is a type of motion blurry in image processing and can be corrected by deconvolution if the vibration kernel is known. In this experiment, we estimated the vibration kernel based on the fact that all atoms should be spherical. First, each experimental image was interpolated by a factor of 2 with linear interpolation, from which a high quality $MoS_2$ region of 100×100 pixels was cropped. Second, a $MoS_2$ atomic model was used to generate reference images at 12 tilt angles. Third, a harmonic oscillator kernel was constructed, of which the vibration direction, amplitude and width were optimized by using the Lucy-Richardson algorithm (*33, 34*) to match the cropped experimental images with the reference ones. Finally, the optimized kernel was applied to the whole experimental image and the image was then binned to its original size. The procedure was repeated for all the 12 experimental images.

Image alignment. Using the x and y coordinates of the W and Mo atoms in the region of interest determined during angle calibration, the center of mass of all W and Mo atoms in each image was aligned to the center of mass of the image with sub-pixel precision.

**3D Reconstruction with sAET**

We implemented sAET reconstruction as described elsewhere (*19*). Briefly, a 3D window of 60×60×60 voxels was chosen and scanned along the x and y axis with a step size of 30 voxels. At each step, the corresponding regions in all 12 images were cropped and grouped into an image stack. Each image stack consists of 12 images with varied shapes, corresponding to the projection of the 3D window along different tilt angles. After a full 2D scan was completed, all the images were partitioned into image stacks. The image stacks were aligned and reconstructed in parallel by the generalized Fourier iterative reconstruction (GENFIRE) algorithm (*18, 35*). Each GENFIRE reconstruction used a 33-voxel support along the z-axis and ran 1000 iterations. To remove these artifacts, we stitched together only the central 30×30×33 voxels of the reconstructed windows to produce a full 3D reconstruction.

**Determination of 3D atomic coordinates and chemical species**

Initial localization of 3D atomic positions and chemical species. The 3D atomic coordinates and chemical species of the 2D heterojunction were traced from the 3D reconstructions with a polynomial fitting method (*31, 37*). The reconstruction was first interpolated onto a finer grid with 3x oversampling using the spline method. All local maxima in the reconstruction were identified and the positions of potential atoms were extracted with a 3D volume (5×5×5 voxels). For every potential atom, a minimum distance of 1.6 Å to its neighbouring atoms was enforced. These positions were manually checked to correct for unidentified or misidentified atoms due to fitting failure or artifacts. We then assigned the initial chemical species based on the $MoS_2$ and $WSe_2$ region and the intensity of the atoms.

Refinement of 3D atomic coordinates. The traced 3D atomic coordinates and chemical species were refined by a gradient descent method as described elsewhere (*18, 19, 31*). The 3D atomic coordinates in each model were refined by minimizing the error between the experimental images and those calculated from the model.

Refinement of chemical species and identification of S/Se vacancies. From the refined 3D atomic coordinates, we applied an atom pair flipping method to identify the S/Se vacancies (*19, 31*), which consists of the following four steps. First, we randomly chose a pair of S/Se atoms between the top and bottom atomic layers. For each selected S/Se pair, projection images were calculated for all the 12 tilt angles by flipping the pair among nine cases: i) both S atoms, ii) both vacancies, iii) both Se atoms, iv) top S and bottom Se atom, v) top Se and bottom S atom, vi) top S atom and bottom vacancy, vii) top vacancy and bottom S atom, viii) top Se atom and bottom vacancy, ix) top vacancy and bottom Se atom. Nine atomic models were generated for the nine cases. Second, nine sets of 12 images with the experimental tilt angles were calculated from the atomic models. An $R_1$ factor was computed between measured and calculated images (*18, 19, 31*). By comparing the $R_1$ factor among the nine cases, the one with the smallest error $R_1$ was chosen and updated in the atomic model. Third, we repeated steps one and two for all the S/Se atom pairs and obtained an updated 3D atomic model. Fourth, we iterated steps one to three for all the S/Se atom pairs until there was no further change in the chemical species.

Dynamic refinement of the S atoms near the W atoms. Due to the use of a low energy electron beam (60 keV) and a small detector inner angle, we observed the dynamic scattering effect of the W atoms, which influenced some nearby S atoms. We identified 19 outlier S atoms violated a minimum distance of 1.6 Å (during refinement) or deviated from the S atomic layer by more than 1 Å. For each outlier S atom, the coordinates of the S atom was scanned with a range of ±40 pm along the x and y axis and ±120 pm along the z axis. The step size was 20 pm, 20 pm and 30 pm along the x, y and z axis, respectively. At each scanning step, we used multislice simulations to calculate 12 images of 60×60 pixels in size at different tilt angles and computed the $R_1$ factor between the calculated and measured images (*18, 19, 31*). The S atom in the atomic model was updated to the position corresponding to the minimum $R_1$ factor. This dynamic refinement procedure was repeated for all outlier S atoms.

**Multislice simulations for 3D precision estimation.**

We performed multislice simulations to estimate the 3D precision of the atomic coordinates. 12 multislice images were computed from the refined atomic model using the same experimental parameters (table S1). Each multislice image was convolved with a Gaussian kernel ($\sigma = 1.1$) to account for the electron probe size, thermal vibrations, and other incoherent effects. The mixture of Gaussian and Poisson noise determined from the experimental images were added to the multislice images. From the 12 multislice images, we used the same imaging processing, sAET reconstruction, atom tracing and refinement procedures to obtain a new 3D atomic model. By comparing it with the experimental 3D atomic model, we found all the chemical species were correctly identified, including all S vacancies. The root-mean-square deviation of the S, Mo, Se and W atoms was estimated to be 15 pm, 4 pm, 15 pm and 6 pm, respectively.

**Measurements of the 3D strain tensor in the $MoS_2$-$WSe_2$ heterojunction**

The strain tensor was determined from the experimental 3D atomic coordinates using a procedure described elsewhere (*19*). The experimental atomic model was separated to $MoS_2$ and $WSe_2$ regions, and each region was aligned to an ideal $MoS_2$-$WSe_2$ atomic model with a least-square fit. The displacement vectors were calculated as the difference in the atomic positions between the experimental and ideal model and interpolated to a cubic grid using a radial basis function (*33*). A 3D Gaussian kernel with $\sigma = 3.16$ Å was convolved with the displacement field to increase the precision of the strain tensor measurement.

**First principles calculations utilizing experimental 3D atomic coordinates without relaxation**

<u>Stitched supercell construction.</u> To investigate the vibrational properties of the lateral heterojunction using first-principles, a periodic supercell across the interface is required. The reconstructed region consists of two crystallographically distinct interfaces with Pmc2 (armchair) and Pmm2 (zigzag) space groups, respectively. We constructed a periodic supercell by stitching together two regions across a zigzag interface. To this end, imagine starting in the $MoS_2$ region and moving perpendicularly towards the middle zigzag interface and into the $WSe_2$ region, stopping at a heavy (W) atom (Fig. 3A, red box). Having moved sufficiently far into the $WSe_2$ region, the effects of the interface on the atomic coordinates should be minimal, and thus we assumed we could equivalently use the displacement of a different W atom, also sufficiently far from the interface. We chose a different W atom and moved perpendicularly towards the bottom zigzag interface into the $MoS_2$ region (Fig. 3A, blue box).

Computational details. We avoided relaxing the experimental coordinates in order to extract vibrational properties by sampling the Born-Oppenheimer (BO) energy surface, around the experimental coordinates local minimum, using *ab initio* molecular dynamics (AIMD). In particular, we used the Vienna *ab initio* simulation package, VASP (*38, 39*) for AIMD, and extract the harmonic force-constants using the ALM code (*40*), as implemented in phonopy (*41*). We used a cut-off energy of 500eV on a Monkhorst-Pack k-points grid of 8x8x1. Phonon dispersion unfolding was performed with the upho code (*21*).

Participation ratios. The frequency-dependent participation ratios and interface-weighted participation ratios were evaluated using,

$$PR_q = \left(\sum_\alpha \frac{|e_\alpha(q)|^2}{M_\alpha}\right)^2 \Big/ N_\alpha \sum_\alpha \frac{|e_\alpha(q)|^4}{M_\alpha^2} \qquad (2)$$

$$IPR_q = \sum_\alpha w_\alpha \frac{|e_\alpha(q)|^2}{M_\alpha} \Big/ \left(N_\alpha \sum_\alpha \frac{|e_\alpha(q)|^4}{M_\alpha^2}\right)^{1/2} \qquad (3)$$

where $e_\alpha(q)$ and $M_\alpha$ are the eigenvectors and atomic masses of species α, respectively (*23*). The interface-weight $w_\alpha$ is a normally-distributed weight of the distance of an atom in the supercell from the interface, normalized to sum up to one interface, thus enabling the 1/144 comparison in the work.

Effect of experimental atomic coordinate relaxation. To capture the influence of the 3D atomic defects on lattice vibrations, the experimental atomic coordinates of the stitched supercell were not relaxed. Figure S4 compares the unfolded phonon dispersions (Fig. 3B) against bulk phonon dispersions and the unfolded phonon dispersion of crystallographic zigzag interface, i.e. one where both sides have been *independently* relaxed to their respective global minimum. While the crystallographic PDOS is more consistent with a linear combination of bulk $MoS_2$ and $WSe_2$, the PDOS directly calculated from the experimental atomic coordinates show localized interface modes, corroborated by spatially resolved EELS (see section below). This study highlights the importance of using the experimental coordinates as direct input to first principles calculations. What is more, the increased density of soft (imaginary-frequency) modes, suggests the out-of-plane ripple and structural reconstructions present in the stitched supercell stabilizes the structure's lattice vibrations (Fig. 3B).

**Spatially resolved vibrational EELS**

The vibrational spectra were measured by the same Nion microscope at 60 kV with a convergence semi-angle of 33 mrad and a probe current of 100 pA, which provides a spatial resolution of 1.5 Å. This higher probe current is beneficial to the acquisition of the vibrational spectra. ADF-STEM images of Fig. 4A and B were acquired with an ADF collection semi-angle ranging from 70 to 210 mrad. With the implementation of an alpha-type monochromator and newly designed EELS spectrometer, the best energy resolution of the EELS is 5.7 meV with a short exposure time of 3 ms. To obtain a sufficient vibrational signal from the 2D heterojunction and avoid the saturation of zero-loss peak (ZLP), an exposure time of 300 ms per frame was used, which corresponds to an energy resolution of 8.3 meV. The EELS collection semi-angle was set to be 20 mrad, which is larger than the size of Brillouin zone of either MoS$_2$ or WSe$_2$ along the [001] direction. The EELS dispersion was 0.44 meV/channel. The high resolution vibrational spectra (Fig. 4) was collected by running our custom-developed python script on top of Nion's Swift software (*24*, *25*). A 40-point line-scan experiment was performed with a step size of 1.6 nm by aligning and summing 200 frames of 300 ms exposures per point. Background was subtracted from the vibrational spectra by fitting a power function to two fitting windows of 16–20 meV and 60–70 meV (fig. S5). A Matlab-based k-means clustering algorithm (*42*) was used to divide the vibrational spectra into three groups (#1–16 for MoS$_2$, #17–26 for interface, and #27–40 for WSe$_2$), where the definition of 'distance' (the minimization parameter) was one minus the correlation between the background-subtracted spectra and the centroid. Each centroid was the component-wise mean of the points in that cluster, after centering and normalizing all spectra to zero mean and unit standard deviation. The process of minimization increased the correlation between each spectrum and its corresponding centroid. For the peak separation of the averaged vibrational spectra, the Levenberg Marquardt iteration algorithm (*43*) was applied to fit spectra by five Gaussian-type peaks without any bounds on all fitting parameters.

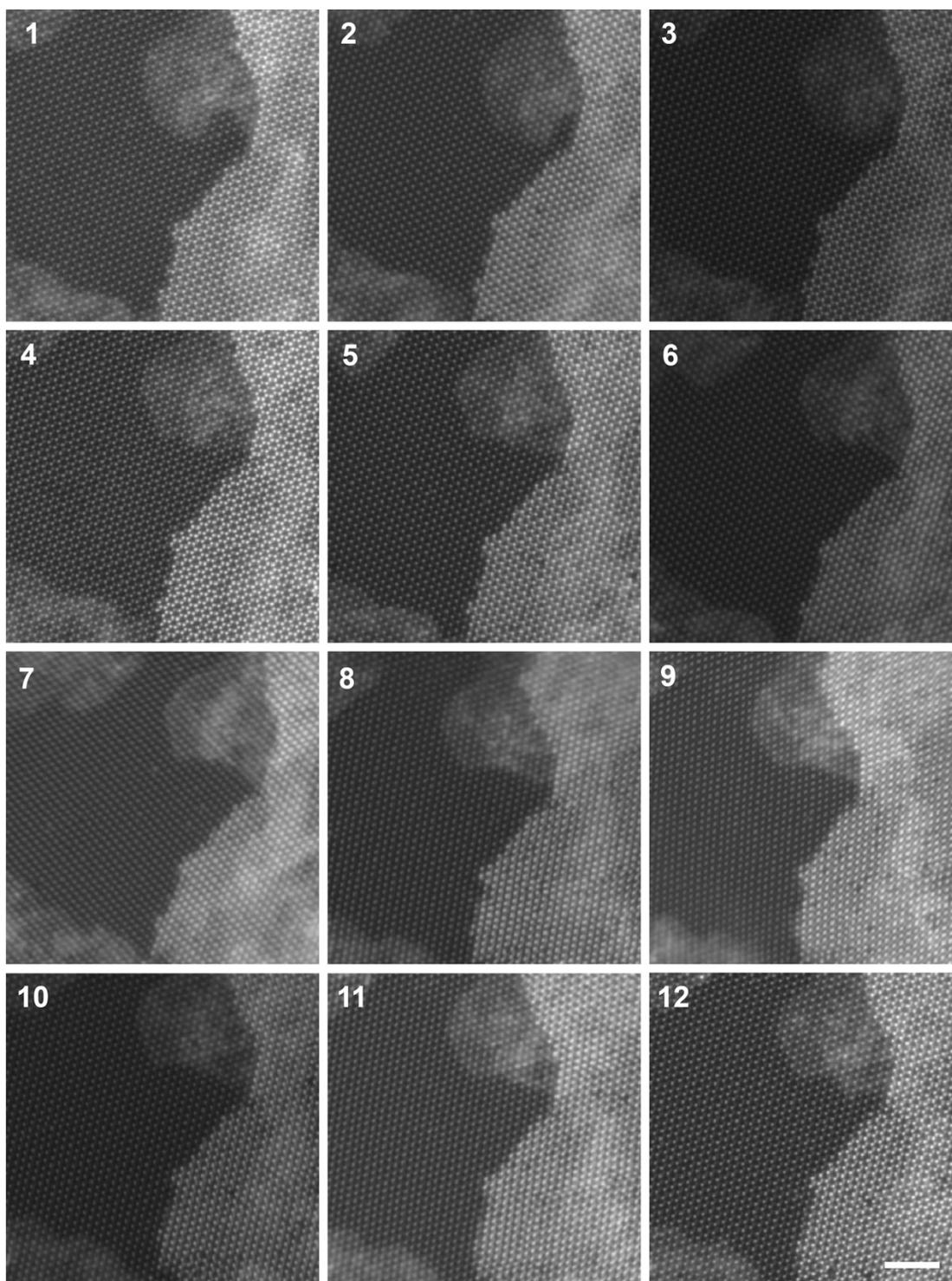

**Fig. S1. Experimental tilt series of the MoS$_2$-WSe$_2$ heterojunction.** The 12 ADF-STEM images after image processing. The tilt angles are shown in table S2. Due to the use of a low electron energy beam and a low-dose acquisition scheme, there was minimal structural change of the heterojunction during the data acquisition. The total electron dose for the whole tilt series was estimated to be $3.8\times10^5$ e/Å$^2$. Scale bar, 2 Å.

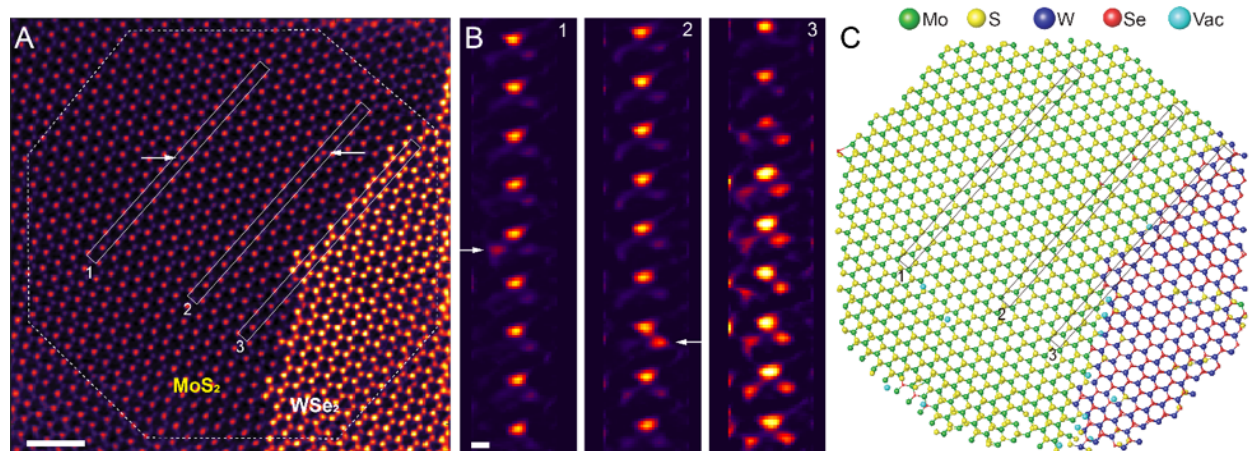

**Fig. S2. 3D reconstruction and experimental atomic model of the MoS$_2$-WSe$_2$ heterojunction.** (**A**) Top view of the reconstruction. The 3D atomic coordinates and chemical species in the dotted octagon were determined with picometer precision. The arrows indicate two Se atoms substitutional of S atoms. (**B**) Side view images of three rectangular regions in (A), where the arrows point to the two substitutional atoms. (**C**) 3D atomic model of the dotted octagon in (A), where the three rectangles show the same regions in (A).

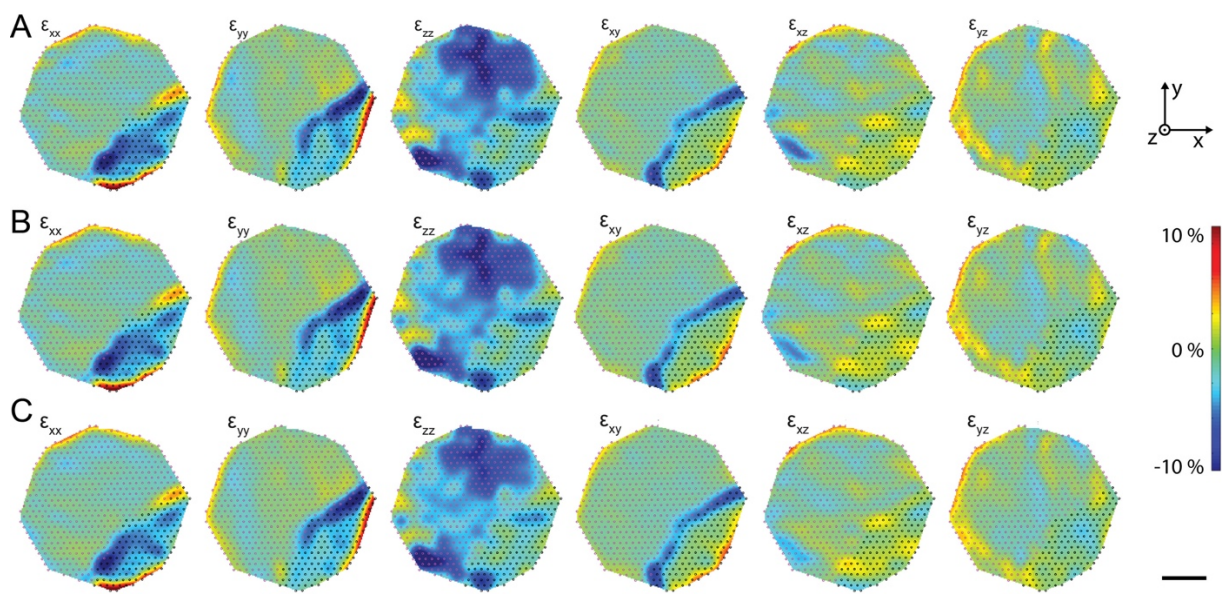

**Fig. S3. 3D strain tensor of the MoS$_2$-WSe$_2$ heterojunction.** Six components of the strain tensor for the top (A), middle (B) and bottom (C) atomic layers.

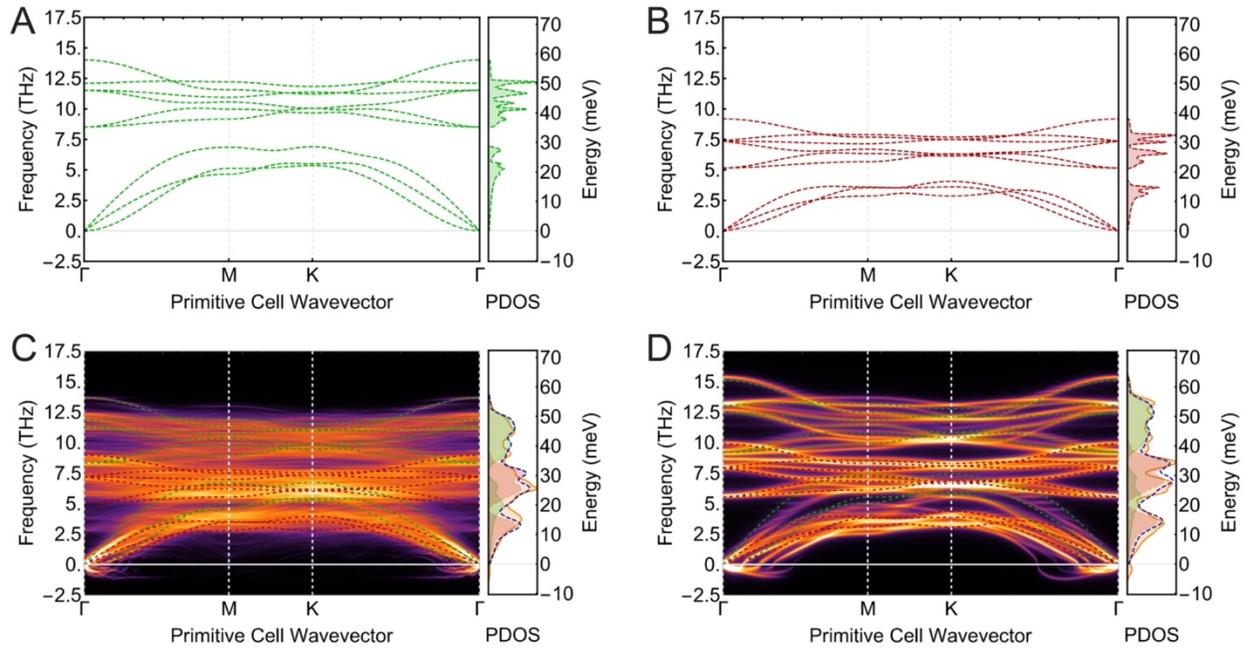

**Fig. S4. Phonon dispersions in experimental and symmetrized interfaces.** Phonon dispersion and corresponding density of states for bulk $MoS_2$ (**A**) and for bulk $WSe_2$ (**B**), following those computed in ref. (*22*). (**C**) Unfolded phonon dispersion computed using the experimental 3D atomic coordinates without relaxation, the same as Fig. 3B of the main text. The phonon dispersions of bulk $MoS_2$ and $WSe_2$ from panels (A) and (B) are overlaid for comparison. In the side panel, the phonon density of states are shown for the interface (orange), along with those for bulk $MoS_2$ (green) and $WSe_2$ (red). The density of states are broadened for ease of comparison. (**D**) Unfolded phonon dispersion and density of states computed for a crystallographic zigzag interface, showing the difference from (C).

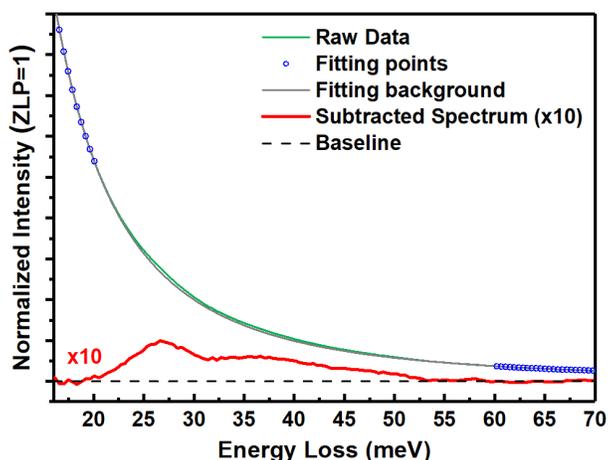

**Fig. S5. Background subtraction of the vibrational spectrum at the interface with two fitting windows.** The background-subtracted spectrum is multiplied by 10 times to improve visualization.

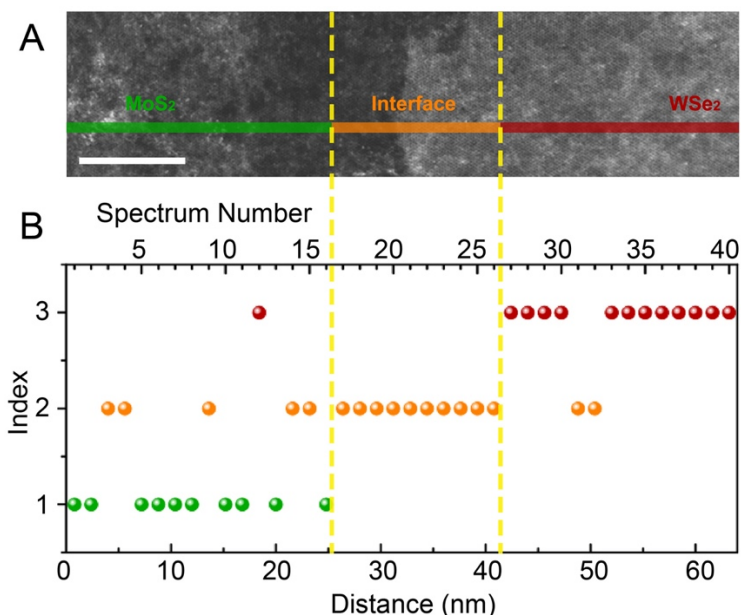

**Fig. S6. K-means clustering analysis of the vibrational spectra in Fig. 4C.** (**A**) Low magnification ADF-STEM image duplicated from Fig. 4A. Scalebar, 10 nm. (**B**) Index of each spectrum after k-means analysis with $k = 3$. The spectra of the $MoS_2$, interface and $WSe_2$ regions were assigned to index 1, 2 and 3, respectively. The slight misassignment was likely due to the noise in the raw data and the interference of the residual polymer during sample preparation. Based on this analysis, we assigned the vibrational spectra #1–16 for the $MoS_2$ region, #17–26 for the interface, and #27–40 for the $WSe_2$ region. The boundaries between these regions are indicated as two yellow dashed lines in (A) and (B). The interface width in (B) is 13 nm, which can be explained by the delocalization of dipole interaction between fast electron and lattice vibrations (*24*, *27*), and is also consistent with other published experimental measurements of both organic and inorganic materials (*44*, *45*).

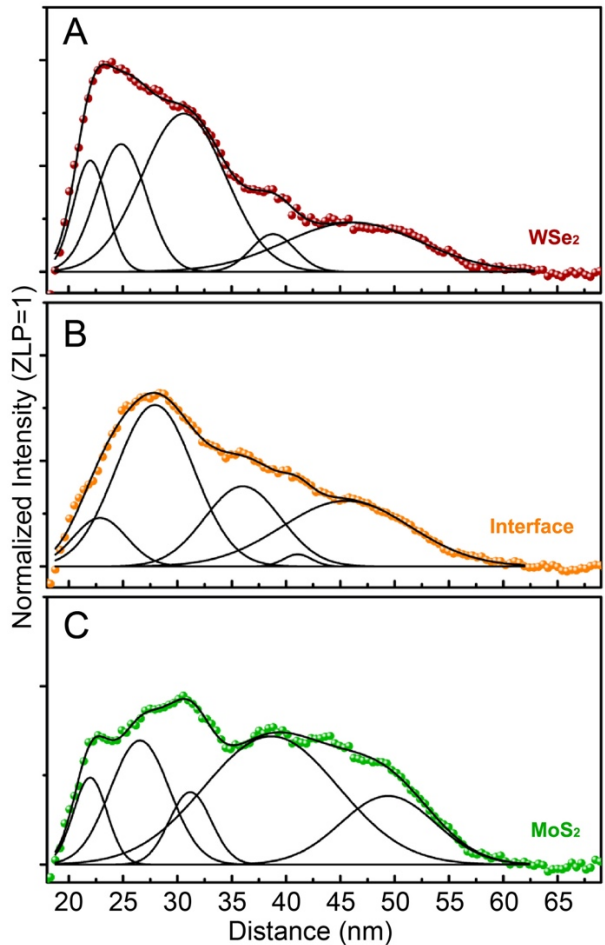

**Fig. S7. Peak fitting of the three averaged vibrational spectra in Fig. 4D.** The peak positions, listed in table S3, were fit with multiple Gaussians. Due to the influence of the ZLP tail, most of acoustic phonon modes below 20 meV could not be measured from our EELS experiment.

**Supplementary Tables**

| Data collection and processing | |
|---|---|
| Voltage (kV) | 60 |
| Convergence semi-angle (mrad) | 30 |
| Probe size (Å) | 1.3 |
| Detector inner angle (mrad) | 30 |
| Detector outer angle (mrad) | 300 |
| Depth of focus (nm) | 10 |
| Pixel size (Å) | 0.32 |
| # of images | 12 |
| Tilt range | See table S2 |
| Electron dose ($10^5$ e/Å$^2$) | 3.8 |
| **Reconstruction** | |
| Algorithm | GENFIRE |
| Interpolation radius (voxel) | 0.3 |
| Oversampling ratio | 3 |
| Number of iterations | 1,000 |
| **Refinement** | |
| $R_1$ (%)$^a$ | 16.0 |
| $R$ (%)$^b$ | 15.0 |
| B' factors (Å$^2$) | |
|    Mo atoms | 13.7 |
|    S atoms | 12.9 |
|    W atoms | 19.0 |
|    Se atoms | 16.3 |
| # of total atoms | 1,886 |
| # of Mo atoms | 488 |
| # of S atoms | 991 |
| # of W atoms | 150 |
| # of Se atoms | 257 |
| # of S/Se vacancies | 16 |

**Table S1. Parameters used for the sAET experiment, image reconstruction and refinement.** $^a$The $R_1$-factor is defined as equation 5 in ref. (*18*). $^b$The $R$ factor was calculated by $R = \frac{\Sigma ||F_{obs}| - |F_{calc}||}{\Sigma |F_{obs}|}$, where $|F_{obs}|$ is the Fourier magnitude obtained from experimental images and $|F_{calc}|$ is the Fourier magnitude calculated from a 3D atomic model.

|  | Nominal angles (°) | | Calibrated angles (°) | | |
| --- | --- | --- | --- | --- | --- |
| Tilt axis | α | β | Z | Y | X |
|  | [0 -1 0] | [-1 0 0] | [0 0 1] | [0 1 0] | [1 0 0] |
| Image #1 | 0 | 0 | 0.3 | 1.1 | -2.0 |
| #2 | -14.3 | -22.9 | 0.2 | -14.5 | 15.9 |
| #3 | -14.3 | 17.1 | 0.4 | -8.3 | 8.4 |
| #4 | 20.1 | 17.1 | 0.3 | -0.7 | 0.1 |
| #5 | 20.1 | -20.1 | -0.3 | 12.5 | -12.2 |
| #6 | 12.6 | -12.6 | -0.6 | 19.5 | -18.4 |
| #7 | 12.6 | 12.6 | -1.0 | 24.2 | -24.8 |
| #8 | -12.6 | 12.6 | 0.0 | 26.5 | 7.8 |
| #9 | -12.6 | -12.6 | 0.2 | 24.5 | 20.7 |
| #10 | -6.3 | -6.3 | 0.3 | 19.8 | 0.7 |
| #11 | -6.3 | 6.3 | 0.1 | 19.9 | 11.8 |
| #12 | 6.3 | 6.3 | 0.4 | 13.8 | 3.6 |

**Table S2**. Angle calibration for a double tilt Nion stage. The nominal angles were read out from the microscope. The angles were calibrated by 3D coordinate fitting of the W and Mo atoms using least square minimization algorithm (see Materials and Methods in detail).

| # | MoS$_2$ (EELS) | MoS$_2$ (DFT) | WSe$_2$ (EELS) | WSe$_2$ (DFT) | Interface (EELS) |
|---|---|---|---|---|---|
| Peak 1 | 49.4±1.0 | 47.16 | 46.2±0.5 | - | 45.5±1.4 |
| Peak 2 | 38.6±0.7 | 39.38 | 38.8±0.4 | - | 41.1±0.4 |
| Peak 3 | 31.2±0.9 | - | 30.6±1.4 | 31.17 | 36.0±2.7 |
| Peak 4 | 26.5±0.6 | 26.56 | 24.8±1.4 | 25.21 | 27.9±1.1 |
| Peak 5 | 22.0±0.2 | 20.27 | 22.0±0.4 | - | 22.9±1.7 |
|  | Undetectable | 13.19 | Undetectable | 13.17 | Undetectable |
|  | Undetectable | - | Undetectable | 8.24 | Undetectable |

**Table S3. Peak positions (meV) of the EELS vibrational spectra and the PDOS of first principle calculations by fitting multiple Gaussians on data in Fig. 4D and 3D.** In the EELS MoS$_2$ spectrum, the peaks at 49.4 meV and 38.6 meV correspond to the optical phonon modes of MoS$_2$, while the peaks at 26.5 and 22.0 meV are attributed to the acoustic phonon modes of MoS$_2$. In the EELS WSe$_2$ spectrum, two prominent peaks at 30.6 and 24.8 meV match with the optical phonon modes of WSe$_2$. The optical phonon modes of MoS$_2$ appear in the EELS WSe$_2$ spectrum, and vice versa. This phenomenon can be explained by the delocalization effect of low-energy-loss excitations. However, the signal intensity of vibrational modes from the local region is much stronger than that of delocalized phonon modes from the counterpart because the delocalization signal exponentially decays as the distance away from the counterpart (*26*). Therefore, the major signature in the EELS spectra still reflects the intrinsic vibrational signal of the local region (*46*). Due to the influence of the ZLP tail, most of phonon modes below 20 meV could not be measured from our EELS experiment and are labeled as "undetectable".